\documentclass[runningheads]{llncs}
\usepackage{fontawesome}  
\usepackage{amsmath}
\usepackage{booktabs}
\usepackage{multirow}
\usepackage[table,xcdraw]{xcolor}
\usepackage{graphicx}
\usepackage{textcomp}
\usepackage{subfig}
\usepackage{caption}
\usepackage[utf8x]{inputenc}
\usepackage{csquotes}
\usepackage{flexisym}
\newcommand{\ignore}[1]{}
\usepackage[hidelinks]{hyperref}
\usepackage[font=scriptsize]{caption}

\usepackage[ruled,vlined,linesnumbered]{algorithm2e}

\usepackage{amssymb}
\usepackage[bottom]{footmisc}
\usepackage{wrapfig}
\usepackage{nicefrac}
\usepackage{bm}
\usepackage{mathrsfs}
\usepackage{xcolor}
\usepackage{hyperref}
\usepackage{relsize}
\usepackage{breqn}

\newcommand{\smallblacksquare}{\scriptsize \blacksquare \normalsize}

\newcommand{\pactivities}{\mathcal{A}}
\newcommand{\mechanism}{\mathcal{M}}

\newcommand{\pcases}{\mathcal{C}}

\newcommand{\ptimes}{\mathcal{T}}

\newcommand{\multiset}{\mathcal{B}}

\newcommand{\RN}[1]{%
	\textup{\uppercase\expandafter{\romannumeral#1}}%
}

\usepackage{mathtools}

\newcommand{\simplelog}{\tilde{L}}

\newcommand{\orcid}[1]{
	\href{https://orcid.org/#1}{\includegraphics[scale=0.4]{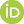}}
}

\begin{document}
%
% \title{Differential Privacy for Privacy-Preserving Continuous Event Data Publishing\thanks{Supported by organization x.}}

\title{Quantifying Temporal Privacy Leakage in Continuous Event Data Publishing\thanks{\scriptsize Funded under the Excellence Strategy of the Federal Government and the L{\"a}nder. We also thank the Alexander von Humboldt Stiftung for supporting our research.}}
\titlerunning{Quantifying Temporal Privacy Leakage in Continuous Event Data Publishing}

\author{Majid Rafiei\inst{1}\orcid{0000-0001-7161-6927}\textsuperscript{\href{mailto:majid.rafiei@pads.rwth-aachen.de}{\faEnvelopeO}} \and
	Gamal Elkoumy\inst{2}\orcid{0000-0002-0491-9722} \and
	Wil M.P. van der Aalst\inst{1}\orcid{0000-0002-0955-6940}}

\authorrunning{Majid Rafiei et al.}

\institute{Chair of Process and Data Science, RWTH Aachen University, Aachen, Germany \\
% \email{\{majid.rafiei,wvdaalst\}@pads.rwth-aachen.de} 
\and
University of Tartu, Tartu, Estonia\\
% \email{gamal.elkoumy@ut.ee}
}
\maketitle              % typeset the header of the contribution
\begin{abstract}
Process mining employs event data extracted from different types of information systems to discover and analyze actual processes. 
Event data often contain highly sensitive information about the people who carry out activities or the people for whom activities are performed.
Therefore, privacy concerns in process mining are receiving increasing attention. To alleviate privacy-related risks, several privacy preservation techniques have been proposed. Differential privacy is one of these techniques which provides strong privacy guarantees. However, the proposed techniques presume that event data are released in only one shot, whereas business processes are continuously executed.
Hence, event data are published repeatedly, resulting in additional risks.
In this paper, we demonstrate that continuously released event data are not independent, and the correlation among different releases can result in privacy degradation when the same differential privacy mechanism is applied to each release. We quantify such privacy degradation in the form of temporal privacy leakages. We apply continuous event data publishing scenarios to real-life event logs to demonstrate privacy leakages.

\end{abstract}
\section{Introduction}\label{sec:introduction}

Process mining forms a family of techniques used to analyze operational processes of organizations. These techniques use event logs extracted from information systems. An event log contains sequences of events, and each event reflects the execution of an activity with some attributes, e.g., the timestamp at which the activity was performed or the case for which the activity was performed. Some event attributes may refer to individuals, e.g., patients or customers, thus raising privacy concerns. 

Data regulations, e.g., GDPR~\cite{GDPR0}, limit the analysis of sensitive event logs. To circumvent such restrictions, Privacy-Preserving Process Mining (PPPM)~\cite{pppm-tmis-short} proposes techniques to guarantee privacy preservation, e.g., \textit{Differential Privacy} (DP)~\cite{dwork2006differential} or \textit{group-based} privacy preservation techniques, i.e., $k$-anonymity and its extensions \cite{RafieiA21Group}. DP works based on a noise injection mechanism that injects noise into published data to ensure that modifying a single user's record in the original data has a small impact on the published data. Such an impact is bounded by $\epsilon$, so-called \textit{privacy budget}. The smaller values of $\epsilon$ result in more noise injection and less privacy leakage.

Process mining techniques, such as \textit{process discovery} and \textit{conformance checking}, discover and analyze the control-flow of a process which is based on the distribution of \textit{trace variants}, i.e., the control-flow aspect of an event log. A trace variant is a sequence of activities performed for a case.
Various privacy mechanisms have been proposed to anonymize the control-flow aspect of an event log~\cite{MannhardtKBWM19_short,RafieiA21Group,SaCoFa,amun}. These approaches consider only a one-shot data release. However, business processes are continuously executed, stressing the need for Continuous Event Data Publishing (CEDP)~\cite{rafiei2021privacy}. 
In CEDP, events are collected up to a certain point in time or meeting a certain condition and published in the form of event logs. This publishing scenario is done continuously based on a \textit{time-window}, e.g., daily, or a \textit{count-window}, e.g., each new release contains one new event per trace.  

% Provided privacy guarantees for every single release of event logs 
% In \cite{rafiei2021privacy}, the authors quantify the privacy 

% \begin{figure}[t]
% 	\centering
% 	\subfloat[\scriptsize Events among traces ]{\includegraphics[width=0.3\textwidth]{figures/example_traces.pdf}\label{fig:example_traces}}
% 	\hfill
% 	\subfloat[\scriptsize Directly-Follows Frequencies
% 	]{\includegraphics[width=0.3\textwidth]{figures/DFR_origin.pdf}\label{fig:example_dfr}} \hfill
% 	\subfloat[\scriptsize Differentially Private Directly-Follows Frequencies 
% 	]{\includegraphics[width=0.3\textwidth]{figures/DFR_DP.pdf}\label{fig:example_dp}} \hfill
	
% 	\caption{Continuous event data release under temporal correlation}
% 	\label{fig:Example}
% \end{figure}

\begin{figure}[t]
	\centering
	\subfloat[\scriptsize An event log containing trace variants with their frequencies, e.g., $trace_1$ happened 4 times. ]{\includegraphics[width=0.34\textwidth]{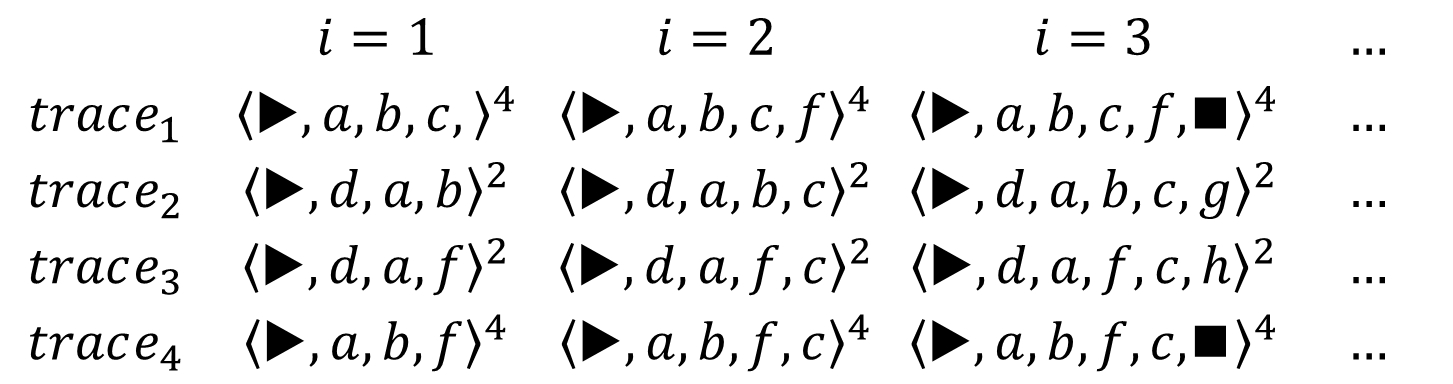}\label{fig:example_traces}}
	\hfill
	\subfloat[\scriptsize The actual frequency of trace variants at each release point.
	]{\includegraphics[width=0.32\textwidth]{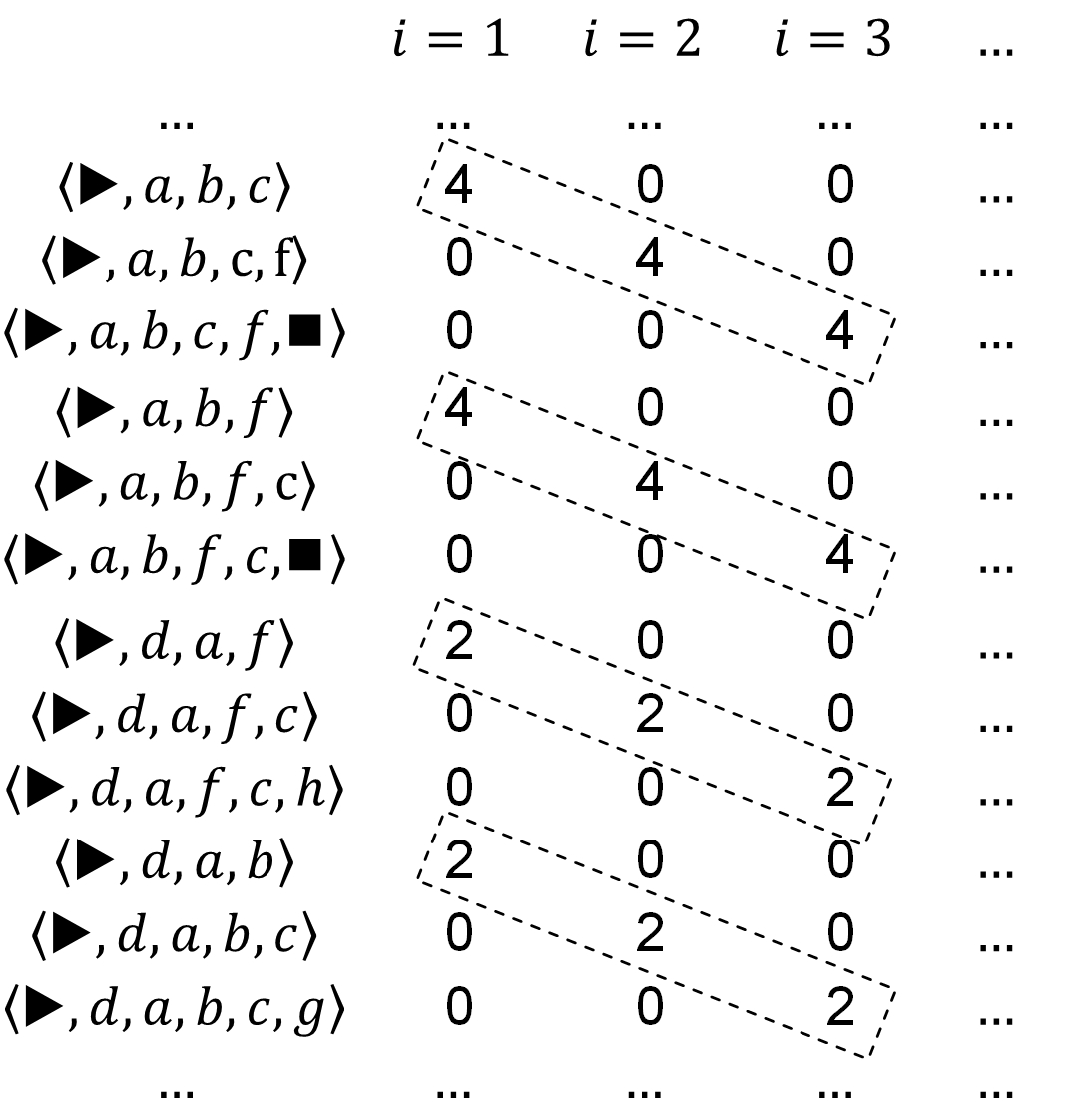}\label{fig:example_traces_original}} \hfill
	\subfloat[\scriptsize Differentially private frequency of trace variants at each release point.     
	]{\includegraphics[width=0.32\textwidth]{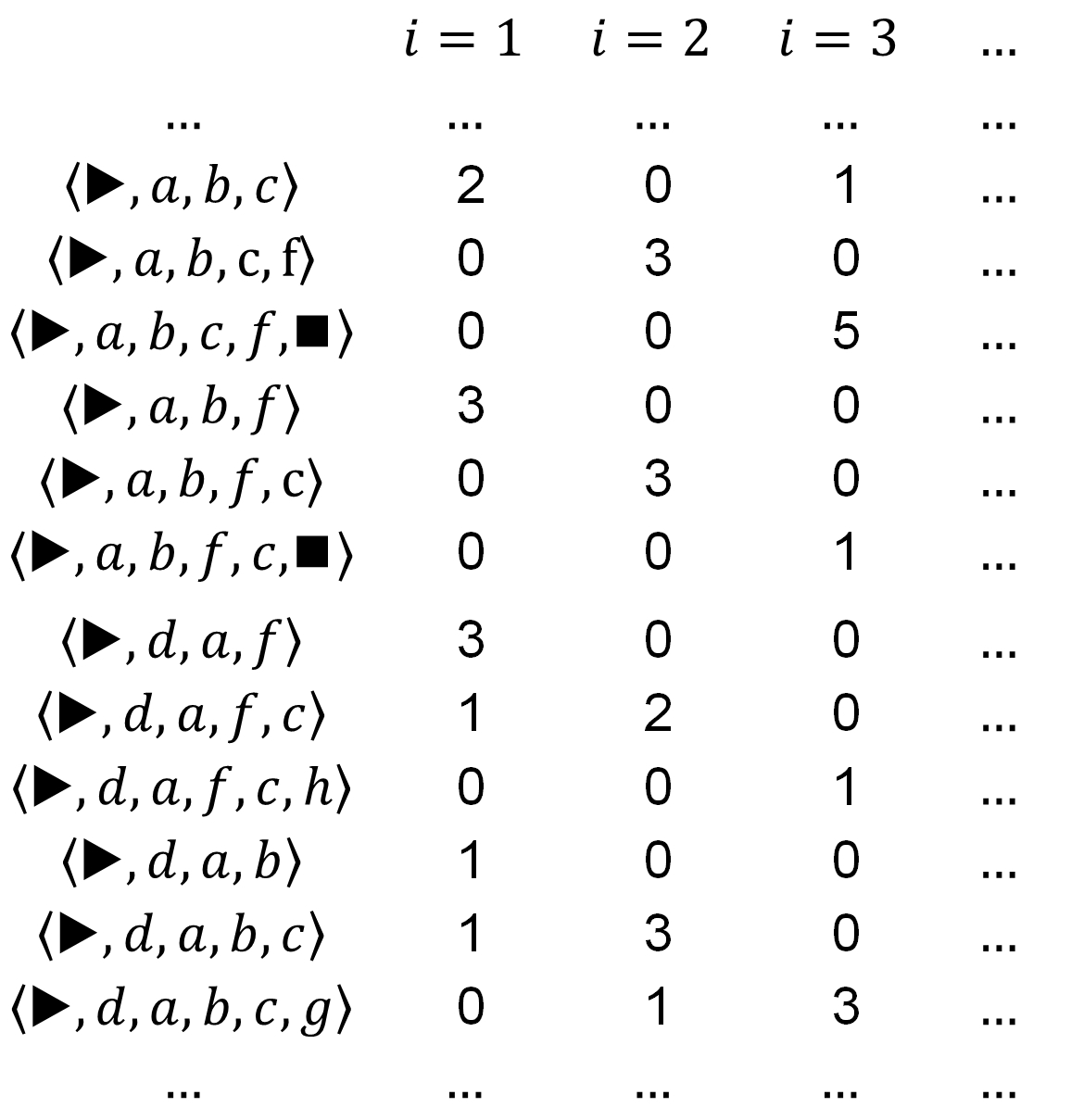}\label{fig:example_traces_dp}} \hfill
	
	\caption{Continuous event data release under temporal correlations.}
	\label{fig:example_cedp}
	\vspace{-0.3cm}
\end{figure}

% CEDP can degrade privacy guarantees provided for separate releases of event logs if there are correlations among continuously released events. For example, consider the continuous release of the event data in Figure~\ref{fig:example_traces} with DP. An organization collects its business process event data and continuously publishes the differentially private directly-follows frequencies of its activities, i.e., the private counts of each two consecutive activities. Suppose that each user goes through only one directly-follows relation at each time point. Adding noise drawn from Laplace mechanism $Lap(1/\epsilon)$ to perturb each frequency~\cite{MannhardtKBWM19_short} achieves $\epsilon$-DP at each time point, as in Fig~\ref{fig:example_dp}. That is because the modification of only one cell in Fig~\ref{fig:example_dfr} affects only one directly-follows relation. However, that may not be true with the existence of temporal correlations. For example, due to the nature of the organization's business process, traces may have a particular subtrace pattern such as ``always going through activity b, then d and e''. Such temporal correlation can be formulated as $Pr(l^t= <d,e> | l^{t-1}=<b,d> )=1$ where $l^t$ is the directly-follows relation of a user at time $t$. In such a case, adding $Lap(1/\epsilon)$ noise only achieves $8\epsilon$-DP because the change of a single directly-follows relation frequency results in a change in 8 different directly-follows frequencies.

CEDP may result in violating the provided privacy guarantees provided for separate releases of event logs if there are correlations among continuously released event logs. For example, consider the continuous release of the event data in Fig.~\ref{fig:example_cedp}. An organization collects its business process event data in the form of trace variants frequencies up to a release point and publishes the differentially private trace variant frequencies. 
Suppose that each case, e.g., a patient, contributes to only one trace variant at each release point and the trace variant of a case is the sensitive information that needs to be protected. 
Note that the trace variant of a case is considered sensitive information because it contains the entire sequence of activities performed for the case. For example, in the healthcare context, the activities are treatment-related, and sequences of activities can be exploited to determine the health conditions of cases, e.g., their diseases.

In order to provide an $\epsilon$-DP guarantee, one needs to hide the participation of an individual in the released output. To this end, the output gets noisified. 
The amount of noise is determined by the privacy parameter $\epsilon$ and the \textit{sensitivity} of a query. The sensitivity indicates how much uncertainty is required to hide the contribution of one individual to the query. Here, the query is the frequency of each trace variant. Since the modification of only one frequency value, i.e., the contribution of one individual, at a specific release point $i$ in Fig.~\ref{fig:example_traces_original} affects only one trace variant, the sensitivity is set to 1.
Adding noise drawn from a \textit{Laplacian distribution} with scale $\nicefrac{1}{\epsilon}$, where the sensitivity value is in the numerator, to perturb each frequency achieves $\epsilon$-DP at each release point \cite{Dwork08differential_short}, as in Fig.~\ref{fig:example_traces_dp}. 

However, that may not be true with the existence of \textit{temporal correlations}. 
For example, as shown in Fig.~\ref{fig:example_traces_original}, the frequency of each trace variant at release point $i$ is not independent of the frequency of its prefix at release point $i{-}1$.  
Thus, adding Laplacian noise with scale $\nicefrac{1}{\epsilon}$ at the release point $i{=}3$ only achieves $3\epsilon$-DP, which is three times weaker than the first provided guarantee. 
One can interpret this situation based on \textit{group differential privacy}, where correlated data are protected as a group \cite{chen2014correlated}. 
Moreover, due to the nature of business processes, traces may have a particular subtrace pattern, such as ``activity b always follows activity a''. Such temporal correlations can be formulated as conditional probabilities to analyze their effect on the provided privacy guarantees by DP mechanisms \cite{QDPContinuous_short}.

In this paper, we adapt the approach introduced in \cite{QDPContinuous_short}. 
In \cite{QDPContinuous_short}, the authors assume that probability matrices explaining the correlations between different releases are given. However, we exploit some characteristics of CEDP to obtain such probabilities. 
We show that different event data publishing scenarios can affect the correlations and the privacy leakage results. 
We also investigate the effect of specific event log characteristics on the correlations and privacy leakages.  
% We use Bayesian inference to quantify the temporal privacy leakage when DP is used for continuous event data publishing. 
Our proposal utilizes a transition system to model traces in the form of states at each point of release. Particularly, we focus on a full-history transition system, so-called prefix automaton, where each state represents a prefix of a trace from the start point until the state. We utilize such a transition system to obtain conditional probabilities between states (traces) at each release point.

The paper is structured as follows. Section~\ref{sec:related_work} discusses related work. Section~\ref{sec:preliminaries} introduces basic notations and formal definitions. Section~\ref{sec:CEDP} demonstrates our approach to quantify temporal privacy leakage in CEDP. In Section~\ref{sec:experiments}, we provide experiments based on real-life public event logs. Section~\ref{sec:conc_disc} concludes the paper and discusses some limitations of the approach.

\section{Related Work}
\label{sec:related_work}

%DP in process mining 
A plethora of studies has been conducted to provide privacy for process mining. In~\cite{pppm-tmis-short}, the authors studied the requirements and challenges of providing privacy-preserving process mining. Several studies applied differential privacy to publish event logs. Mannhardt et al.~\cite{MannhardtKBWM19_short} applied differential privacy to anonymize queries over event logs. PRIPEL~\cite{pripel_short} applies differential privacy to anonymize timestamps of event logs. SaCoFa~\cite{SaCoFa} integrates differential privacy with event log semantics to anonymize the control flow of event logs. In~\cite{amun}, the authors applied differential privacy to event logs in order to prevent singling out individuals using the prefixes/suffixes of their traces. However, all of the above mechanisms assume one-shot data publishing.

%DP with event data
Dwork et al. first studied differential privacy under continual observation, and they presented user-level~\cite{Dworkuserlevel} and event level~\cite{Dworkeventlevel} privacy. Several studies have investigated applying differential privacy in continuous data publishing. Kellaris et al.~\cite{KellarisPXP14} studied the problem of infinite sequences. Fan et al.~\cite{FanXS13} studied differential privacy with a real-time publishing setting. Cao et al.~\cite{QDPContinuous_short} quantified the risk of using differential privacy under temporal correlation to release continuous location data. 
% We adapt the same approach to quantify temporal privacy leakage in CEDP.
A framework for quantifying risk when publishing only one event log for process mining has been studied in~\cite{rafiei_quantification}. 
To the best of our knowledge, no study has presented a risk quantification for differential privacy in continuous event data publishing. Although, in \cite{rafiei2021privacy}, the authors have elaborated possible attacks against continuous anonymized event data publishing, that work focuses on group-based privacy preservation techniques. 
% However, this work focuses on the risk quantification of differential privacy mechanisms.

%Group based and other PPPM
% There are also other studies regarding privacy preservation in process mining that adapted group-based privacy techniques to anonymize event logs \cite{RafieiA21Group,pretsaICPM2019_short}. 
% In \cite{rafieiWA18_short,rafieiWA19_short}, the authors propose an encryption-based framework for providing confidentiality in process mining. 
% Kabierski et al.~\cite{KabierskiFW21} provided privacy-aware process performance indicators. 

% Moreover, Risk quantification of publishing event logs for process mining has been studied in~\cite{VoigtFJKTMLW20,rafiei_quantification}. 

% Process mining based on distributed event logs in the inter-organizational setting has been studied in~\cite{ElkoumyFDLPW20}. 
% A privacy extension for XES files has been proposed in~\cite{RafieiA20}. 
% Some tools also provided for supporting and integrating the proposed peivacy preservation techniques in process mining, e.g., PC4PM \cite{rafiei_PC4PM_short}.

\section{Preliminaries}\label{sec:preliminaries}

In this section, we provide formal definitions for \textit{event logs}, \textit{transition systems}, and \textit{differential privacy}, which will be used to explain the approach.  

% For $\sigma_1, \sigma_2 \in A^*$, $\sigma_1 \sqsubseteq \sigma_2$ if $\sigma_1$ is a subsequence of $\sigma_2$, e.g., $\langle a,b,c,x \rangle \sqsubseteq \langle z,x,a,b,b,c,a,b,c,x \rangle$. 
% For $\sigma \in A^*$, $\{a \in \sigma\}$ is the set of elements in $\sigma$, and $[a \in \sigma]$ is the multiset of elements in $\sigma$, e.g., $[a \in \langle x,y,z,x,y \rangle ] = [x^2,y^2,z]$.
% For $x=(a_1,a_2,...,a_n) \in A_1 {\times} A_2 {\times} ... {\times} A_n$, where $ A_1 {\cap} A_2 {\cap} ... {\cap} A_n {=} \emptyset$, $\pi_{A_i}(x) {=} a_i$ is the projection of the tuple $x$ on the element from the domain $A_i$, $1{\le} i {\le} n$. 

\subsection{Event Log}\label{subsec:event_log}
For a given set $A$, $A^*$ is the set of all finite sequences over $A$, and $\multiset(A)$ is the set of all multisets over the set $A$. 
% For $A_1,A_2 \in \multiset(A)$, $A_1 \subseteq A_2$ if for all $a \in A$, $A_1(a) \leq A_2(a)$.
A finite sequence over $A$ of length $n$ is a mapping $\sigma \in \{1,...,n\} \rightarrow{A}$, represented as $\sigma = \langle a_1,a_2,...,a_n \rangle$ where $a_i = \sigma(i)$ for any $1\leq i \leq n$. $|\sigma|$ denotes the length of the sequence.
Given $A$ and $B$ as two multisets, $A \uplus B$ is the sum over multisets, e.g., $[a^2,b^3] \uplus [b^2,c^2] = [a^2,b^5,c^2]$.
A multiset set $A$ can be represented as a set of tuples $\{ (a,A(a)) | a \in A \}$ where $A(a)$ is the frequency of $a \in A$.
For $\sigma_1, \sigma_2 {\in} A^*$, $\sigma_1 {\sqsubset} \sigma_2$ if $\sigma_1$ is a subsequence of $\sigma_2$, e.g., $\langle z,x,a,b \rangle {\sqsubset} \langle z,x,a,b,b,c,a,b,c,x \rangle$.

\begin{definition}[Event]
	\label{def:event}
	An event is a tuple $e {=} (c,a,t)$, where $c {\in} \pcases$ is the \textit{case identifier}, $a {\in} \pactivities$ is the activity associated with the event $e$, and $t {\in} \ptimes$ is the timestamp of the event $e$.
	We call $\xi {=} \pcases {\times} \pactivities {\times} \ptimes$ the universe of events. 
	Given an event $e=(c,a,t) \in \xi$, $\pi_{case}(e)=c$, $\pi_{act}(e)=a$, and $\pi_{time}(e)=t$.
\end{definition}

Note that $\blacktriangleright$ and $\blacksquare$ are artificial start and end activities included in $\pactivities$, i.e., $\{ \blacktriangleright,\blacksquare\} {\subset} \pactivities$. 
We assume that the case identifiers are dummy identifiers referring to individuals such as patients, workers, customers, etc. These identifiers cannot be exploited to directly re-identify individuals.

\begin{definition}[Trace, Trace Variant]
	\label{def:trace}
	Let $\xi$ be the universe of events. A trace $\sigma{=}\langle e_1,e_2,...,e_n \rangle$ in an event log is a sequence of events, s.t., for each $e_i,e_j {\in} \sigma$, $1{\leq}i{<}j{\leq}n$: $\pi_{case}(e_i){=}\pi_{case}(e_j)$, and $\pi_{time}(e_i) {\le} \pi_{time}(e_j)$.
	A \textit{trace variant} is a trace where all the events are projected on the activity attribute, i.e., $\sigma {\in} \pactivities^*$.
\end{definition}

\begin{definition}[Event Log]
	\label{def:eventlog}
	An \textit{event log} $L$ is a set of case identifiers and their corresponding trace variants, i.e., $L {\subseteq} \pcases {\times} \pactivities^*$. If $(c_1,\sigma_1)$,$(c_2,\sigma_2) {\in} L$ and $c_1{=}c_2$, then $\sigma_1{=}\sigma_2$.
	$\tilde{L} {=} [\sigma \mid (c,\sigma) {\in} L]$ is the multiset representation of traces in $L$, i.e., $\tilde{L} \in \multiset(\pactivities^*)$.
	Given $(c,\sigma) {\in} L$, $\pi_{case}((c,\sigma)){=}c$ and $\pi_{trace}((c,\sigma)){=}\sigma$.
% 	We denote $\universe_{L}$ as the universe of event logs.
\end{definition}

For instance, $L_1=[
(c_1,\langle \blacktriangleright,a,b,c,f,\smallblacksquare \rangle),
(c_2,\langle \blacktriangleright,a,b,f,c,\smallblacksquare \rangle),
(c_3,\langle \blacktriangleright,d,a,b,c,\\g \rangle),
(c_4,\langle \blacktriangleright,d,a,f,c,h \rangle)]$ is an event log with artificial start activities for all the traces, and artificial end activities for the \textit{complete traces}, i.e., the traces of $c_1$ and $c_2$. The traces of $c_3$ and $c_4$ are called \textit{partial traces}, i.e., traces that have not yet reached the end activity.
Note that our definition of an event log represent the control-flow perspective that is the focus of this work. In general, events of an event log may contain more attributes, e.g., \textit{resources}, who perform activities. 

% In the following, We define $state_{hd}()$ and $state_{tl}()$ as the state representation functions describing the current state by the history and the future of the case, respectively.
% $state_{tl}(\sigma,k)=\langle a_{k+1},a_{k+2},...,a_n \rangle$ is a state representation function that considers the last $k$ events of $\sigma$ as a state.

\subsection{Transition System}\label{subsec:transition_system}
\begin{figure}[t]
\centerline{\includegraphics[width=0.75\columnwidth]{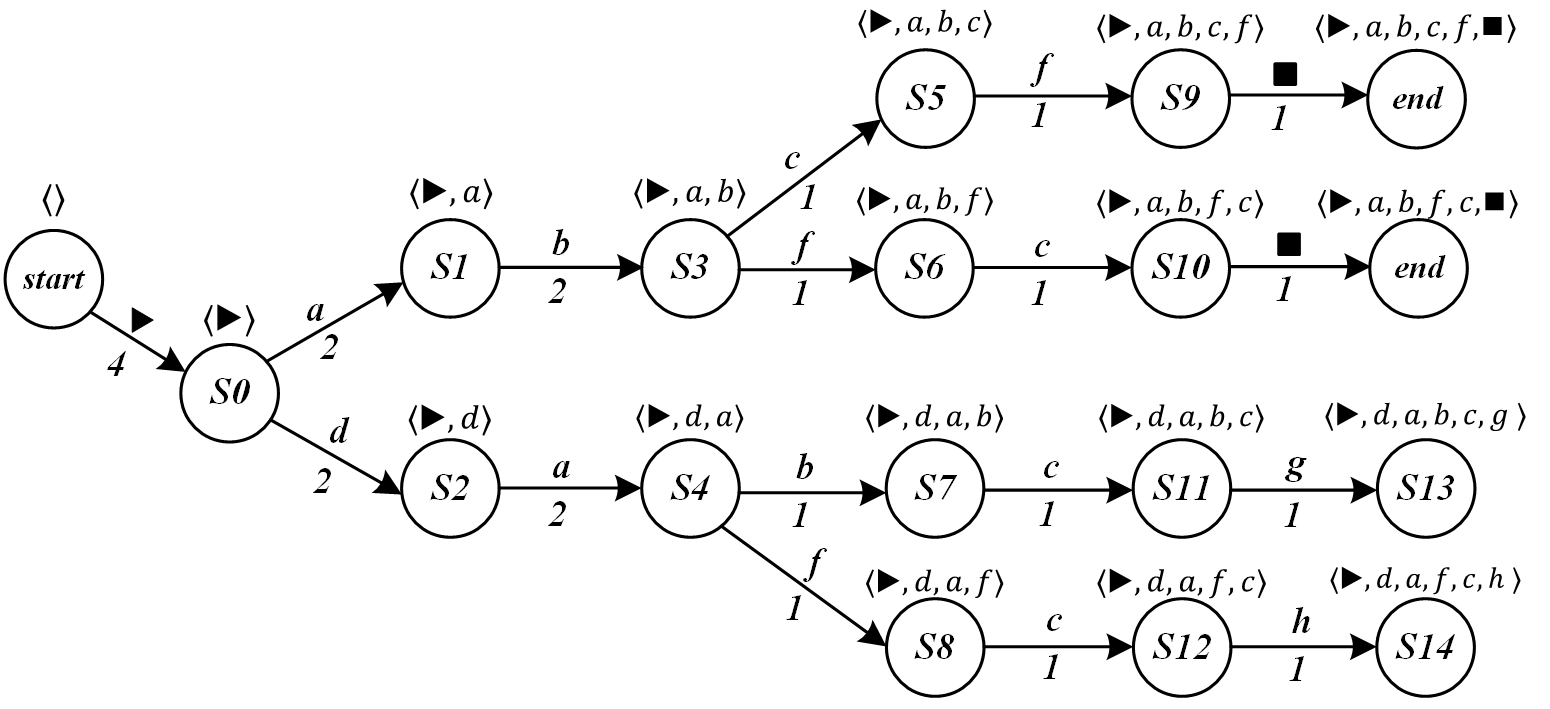}}
\caption{The history transition system of the event log $L_1$. The circles represent states, and the arcs represent transitions with activity names as their labels. The numbers below arcs show the frequency of the corresponding transition.}
	\label{fig:TS_history}
	\vspace{-0.3cm}
\end{figure}

In this paper, we aim to quantify the privacy degradation in CEDP due to the correlations among event logs in different release points. To this end, we need to adopt an event log representation that helps to study these correlations. We consider a full-history transition system, so-called prefix automaton, as the event log representation.
A transition system is one of the most basic process modeling notations which consists of states and transitions. States are represented by circles having unique labels, and transitions are represented by directed arcs with activity labels. Each transition connects two states. Figure~\ref{fig:TS_history} shows a transition system for the event log $L_1$. The labels of states are specified by a state representation function, which is defined as follows.  

\begin{definition}[State]
	\label{def:state}
	 Given $\sigma {\in} \pactivities^*$ as a trace and $0 {\leq} k {\leq} |\sigma|$ as a number, which indicates the number of events of $\sigma$ that have occurred, $state(\sigma,k)$ is a function that produces a state.
\end{definition}

We define $state_{hd}()$ as the state representation functions describing the current state by the history of the case, i.e., given $\sigma=\langle a_1,a_2,...,a_n \rangle$ as a trace of length $n$, $state_{hd}(\sigma,k)=\langle a_1,a_2,...,a_k \rangle$.

\begin{definition}[Event Log Representation]
	\label{def:ts}
    Let $L {\subseteq} \pcases {\times} \pactivities^*$ be an event log and $state()$ be a state representation function. $TS_{L,state()}=(S,A,T)$ is a transition system that represents $L$ based on $state()$ where:
    \begin{itemize} 
        \footnotesize
        \item $S=\{ state(\sigma,k) \mid (c,\sigma) {\in} L \wedge 0 {\leq} k {\leq} |\sigma| \}$ is the state space;
        \item $A=\{ \sigma(k) \mid (c,\sigma) {\in} L \wedge 1 {\leq} k {\leq} |\sigma| \}$ is the set of activities;
        \item $T=[ (state(\sigma,k),\sigma(k+1),state(\sigma,k{+}1)) \mid (c,\sigma) {\in} L \wedge 0 {\leq} k {<} |\sigma| ]$ is the multiset of transitions;
        \item $S^{start}=\{ state(\sigma,0) \mid (c,\sigma) {\in} L \}$ is the set of start states; and
        \item $S^{end}=\{ state(\sigma,|\sigma|) \mid (c,\sigma) {\in} L \wedge \sigma(|\sigma|)=\blacksquare \}$ is the set of end states.
    \end{itemize}
\end{definition}

% Using $state_{hd}()$ and $state_{tl}()$ as state representation functions, we create two transition systems where states are representing \textit{history} and \textit{future}, respectively.  

Using $state_{hd}()$ as a state representation function, one can create a transition system where states represent prefixes. 
Consider $L_1=[
(c_1,\langle \blacktriangleright,a,b,c,f,\smallblacksquare \rangle),
(c_2,\langle \blacktriangleright,a,b,f,c,\smallblacksquare \rangle),
(c_3,\langle \blacktriangleright,d,a,b,c,g \rangle),
(c_4,\langle \blacktriangleright,d,a,f,c,h \rangle) ]$ as an event log where $c_1$ and $c_2$ have complete traces, and $c_3$ and $c_4$ have partial traces. 
Figure~\ref{fig:TS_history} shows the history transition system, obtained by considering $state_{hd}()$ as the state representation function for the event log $L_1$.
A history transition system can be converted to a probabilistic model to show the correlation between states as conditional probabilities. We utilize such representation of an event log to quantify the correlations between traces. Then, such correlations are used to quantify temporal privacy leakages of a DP mechanism in CEDP. 

\subsection{Differential Privacy}\label{subsec:differential_privacy}
Differential privacy provides a formal definition of data privacy. 
The main idea of differential privacy is to randomize the data in such a way that an observer seeing the randomized output cannot tell if a specific individual's information was used in the computation \cite{Dwork08differential_short}.
Considering the distribution of trace variants as our sensitive event data, $\epsilon$-DP can be defined as follows.

% An algorithm which is used to randomize data to provide differential privacy is called a \textit{randomized mechanism}.

% \begin{definition}[$\epsilon$-DP]
% 	\label{def:DP}
% 	Let $\mechanism$ be a randomized mechanism with $rng(\mechanism)$ as the set of its possible outputs, and let $D$ and $D'$ be two neighboring datasets, i.e., they differ on a single data point. $\mechanism$, which randomizes a given dataset, satisfies $\epsilon$-differential privacy if for all $R {\subseteq} rng(\mechanism)$ and for any pair of $D,D'$:
% 	\footnotesize
% 	\[
% 	    log\frac{Pr(\mechanism(D) {\in} R)}{Pr(\mechanism(D') {\in} R)} \le \epsilon
% 	\]
% 	\normalsize
% \end{definition}

\begin{definition}[$\epsilon$-DP]\label{def:DP}
    Let $L_1$ and $L_2$ be two neighbouring event logs that differ only in a single entry, e.g., $\tilde{L}_2 {=} \tilde{L}_1 \uplus [\sigma]$, for any $\sigma {\in} \pactivities^*$, and let $\epsilon \in \mathbb{R}_{>0}$ be the privacy parameter.
    A randomized mechanism $\mechanism{:} \multiset(\pactivities^*) {\to} \multiset(\pactivities^*)$ provides $\epsilon$-DP if for any $(\sigma,f) \in \pactivities^* {\times} \mathbb{N}_{>0}$ and for all $\simplelog^' \in rng(\mechanism)$:
    \footnotesize
    \[
     log\frac{Pr((\sigma,f) \in \simplelog^' \mid \mechanism(\tilde{L}_1))}{Pr((\sigma,f) \in \simplelog^' \mid \mechanism(\tilde{L}_2))} \leq \epsilon
    \]
    \normalsize
\end{definition}

The parameter $\epsilon$ is called the \textit{privacy budget} and represents the degree of privacy. The smaller the privacy budget, the stronger the privacy guarantees.
A real-valued query $q$ can be made differentially private by using a \textit{Laplace mechanism} where the noise is drawn from a \textit{Laplacian distribution} with scale $\nicefrac{\Delta q}{\epsilon}$. $\Delta q$ is called the sensitivity of the query $q$. 
Intuitively, $\Delta q$ denotes the amount of uncertainty that one needs to incorporate into the output to hide the contribution of single occurrences at the $\epsilon$-DP level. 
In our context, $q$ is the frequency of a trace variant. Since one individual, i.e., a case, contributes to only one trace, the sensitivity is $\Delta q {=} 1$ \cite{MannhardtKBWM19_short,SaCoFa}. If an individual can appear in more than one trace, the sensitivity needs to be accordingly increased assuming the same value for privacy parameter $\epsilon$ \cite{Dwork08differential_short}.

% \begin{definition}[Laplace Mechanism]
% Let $Q{:} D {\to} {\mathbb{R}}$ be a statistical query on a database $D$, and let $D$ and $D'$ be two neighboring datasets. Given $\Delta Q$ as the maximum L1 norm between $Q(D)$ and $Q(D')$, $\epsilon$-DP can be achieved by adding noise drawn from Laplacian noise with scale $\nicefrac{\Delta Q}{\epsilon}$ to the results of query $Q$.
% \end{definition}

\section{Continuous Event Data Publishing}\label{sec:CEDP}

\begin{figure}[t]
\centerline{\includegraphics[width=0.95\columnwidth]{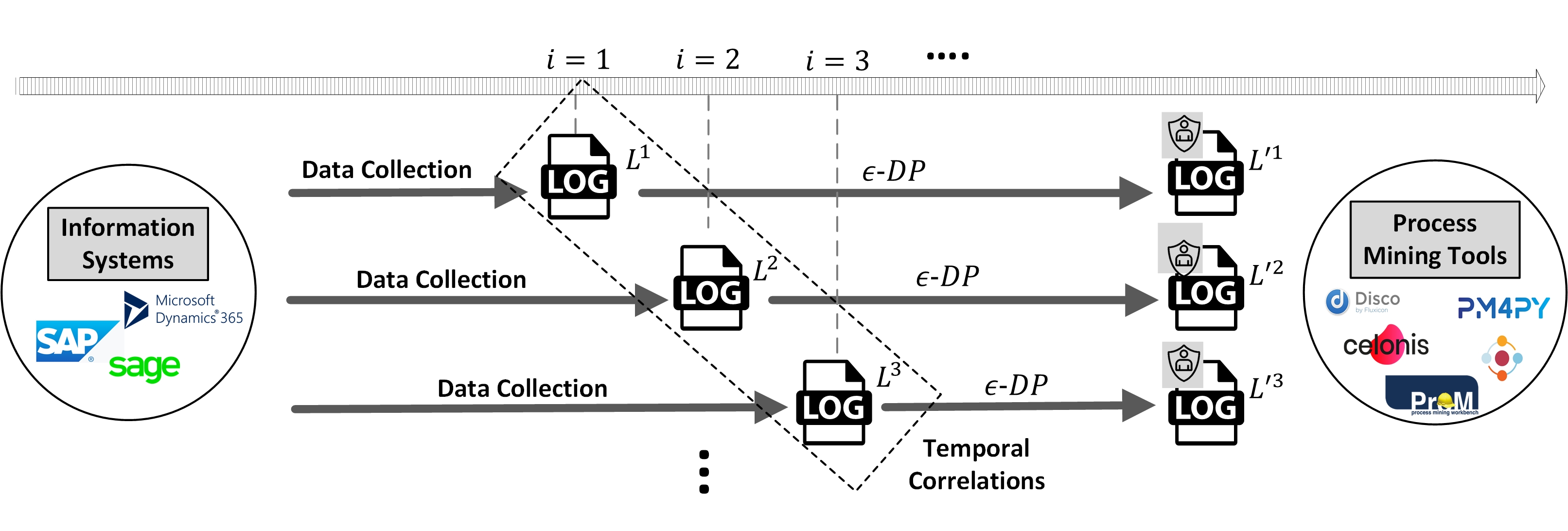}}
\caption{The general overview of continuous event data publishing in process mining.}
	\label{fig:cedp_overview}
	\vspace{-0.3cm}
\end{figure}

Continuous data publishing can generally be classified into three main categories: \textit{incremental}, \textit{decremental}, and \textit{dynamic} \cite{fung2010introduction}. In incremental continuous data publishing, the raw data are cumulatively collected up to a release point, and they cannot be updated or deleted after the collection phase. In decremental continuous data publishing, the previously collected raw data can only be deleted in the later releases. Dynamic data publishing assumes that new raw data can be added to the previously collected data, and the previously collected data can be updated or deleted. 
In the context of process mining, the events generated by an information system are cumulatively collected, and they are not updated or deleted after generation, i.e., the continuous event data publishing is \textit{incremental}. 
Figure~\ref{fig:cedp_overview} shows the general overview of continuous event data publishing using an $\epsilon$-DP mechanism in process mining. Events recorded by information systems are collected up to a release point $i$, then $\epsilon$-DP mechanisms are applied to provide privacy guarantees for each event log $L^i$.

The incremental nature of CEDP can be considered as the main reason of temporal correlations among event logs that need to be published at different release points. For example, the complete traces in an event log $L^i$ appear in all the next releases $L^{i+1},L^{i+2},\cdots$. Moreover, each trace $\sigma$ in an event log $L^i$ has a prefix in all the previous releases $L^j, L^{j+1}, \cdots, L^{i-1}$, s.t., $j<i$ and $L^j$ is the event log where the process of the case having the trace $\sigma$ started. As these examples show, temporal correlations can be categorized into two main categories: \textit{forward} and \textit{backward}. Given $L^i$ as an event log at release point $i$, the former considers temporal correlations between $L^i$ and its next releases, and the latter concerns temporal correlations between $L^i$ and its previous releases. 

\subsection{CEDP Scenarios}\label{subsec:scenarios}
Different event data publishing scenarios can have a significant impact on the privacy leakage based on temporal correlations. In the following, we briefly explain some of the different possible scenarios. 
In general, CEDP scenarios can be based on a \textit{time-window}, e.g., weekly, or a \textit{count-window}, e.g., the number of new cases. 
Since time-window-based scenarios are not deterministic in terms of the amount of new data that can be published in each window, we focus on count-window-based scenarios to quantify the potential privacy degradation.
% Note that time-window based scenarios can be interpreted as count-window based scenarios. For example, by comparing two event logs collected up to the end of first week and second week, one can see the count-based differences, e.g., the number of new events per trace in the second release. 
One can consider different count-window-based scenarios. For example, an event log is released when there exist $x$ new cases compared to the previous release, or when there exist $x$ new events per trace, or when there exist up to $x$ new events per trace, etc. 
We classify the count-window-based scenarios into two main types: \textit{certain} and \textit{uncertain}. The former specifies an exact number, e.g., $x$ new events per trace. The latter specifies a bound, e.g., up to $x$ events per trace.
This classification allows us to assess the effects of certain and uncertain CEDP scenarios on temporal privacy leakages. 

Since events are the smallest units of event logs, to propose a generic approach, we consider the following certain and uncertain scenarios: (S1) an event log is released when there exist exactly $x$ new events per trace compared to the previous release, and (S2) an event log is released when there exist up to $x$ new events per trace compared to the previous release. 
In practice, such bounds can be specified to keep the process mining findings updated.
Note that in both scenarios, events can belong to a new case or an existing one. 
In Subsection~\ref{subsec:leakage_con}, we demonstrate how to use transition systems to quantify the forward and backward privacy leakages considering these scenarios.

\vspace{-0.15cm}
\subsection{Notation Summary}\label{subsec:notations}
For the sake of simplicity, we assume that the number of releases is $\lambda$, which does not need to be exactly specified.  
For a given event log $L$, $C_L {\subset} \pcases$ and $A_L {\subset} \pactivities$ are considered as the finite set of dummy case identifiers and the set of activities that can appear in different releases of $L$, respectively. 
% We assume that all the releases share the same set of cases, and the set of activities is given. 
$L^i$ denotes an event log that needs to be released at point $i {\in} [1,\lambda]$.
$L^i$ contains cases and their current states describing \textit{full history}, i.e., traces. 
% For example, the last states in Figure~\ref{fig:TS_history} represent the traces belonging to the cases in the event log $L_1$ (see Definition~\ref{def:ts}).
% $\tilde{L}^i$ denotes the multiset representation of traces in $L^i$. 
$\sigma_c^i {\in} \Tilde{L}^i$ is the state of a case $c$ at the release point $i$.
Note that according to Definition~\ref{def:eventlog}, each case can only have one trace in an event log.

We consider $\mechanism^i$ as the DP mechanism, which is applied to $\tilde{L}^i$ to randomize the count of trace variants. $rng(\mechanism^i)$ denotes the set of all possible outputs that $\mechanism^i$ can produce. 
For simplicity, $\mechanism^i$ is considered to be the same DP mechanism, e.g., a Laplace mechanism, but maybe with different privacy budgets at each $i {\in} [1,\lambda]$.
% A $\mechanism^i$ is a Laplacian mechanism with a privacy budget of $\epsilon_i$ for each $i \in [1,\lambda]$.
$\simplelog^{'i} {\in} rng(\mechanism^i)$ denotes a differentially private output at release point $i$.
In the following, we first demonstrate the potential privacy loss of $\mechanism^i$ for a single release of event log at release point $i$. Then, we quantify the privacy leakage in the context of continuous releases when $i$ varies from 1 to $\lambda$.  

% We assume that in each release, one activity per each partial trace is released, i.e., $\lambda {\geq} max_{(c,\sigma) \in L}|\sigma|$ for releasing the entire event log.

% In the context of continuous event data release, we define two different privacy goals: \textit{state-level} and \textit{user-level}.
% The former protects each case's single state at release point $i$, i.e., $\sigma_c^i$.
% The later protects all states of each case in the release range, i.e., case's complete trace.

\subsection{Privacy Leakage of a Single Release}\label{subsec:leakage_single}
Consider an adversary whose target is to identify the state of a case $c \in C_L$ at release point $i \in [1,\lambda]$. We assume that such an adversary has the knowledge of all the states at the given release point except the state of the target case $c$.

\begin{definition}[Adversary without Temporal Correlations - $Ad^{L^i_c}$]
	\label{def:bk}
	Let $L^i$ be an event log that needs to be released at the point $i$ and $\sigma_c^i {\in} \tilde{L}^i$ be the state of case $c$ at the point $i$. $Ad^{L^i_c}$ denotes an adversary whose target is to identify $\sigma_c^i$. $L^i_c{=}\{(c',\sigma) {\in} L^i {\mid} c {\neq} c' \}$ is the background knowledge of such an adversary.
\end{definition}

% \begin{definition}[Background Knowledge]
% 	\label{def:bk}
% 	Let $L^i$ be an event log that needs to be released at release point $i$, $s_c^i$ be the state of a case $c$ at the point $i$, and $Ad_c^i$ be an adversary whose target is to identify $s_c^i$. The background knowledge of the adversary is $bk^{L^i}_{s_c^i}=L^i\setminus \{(c,\sigma) {\in} L^i \mid s_c^i{=}\pi_{\sigma}((c,\sigma)) \}$.
% \end{definition}

$Ad^{L^i_c}$ observes $\simplelog^{'i} {\in} rng(\mechanism^i)$ and tries to distinguish case $c$'s state. 
The privacy leakage of the DP mechanism $\mechanism^i$ can be formulated as follows, where $\sigma_c^i,{\sigma'}_c^i {\in} A_L^*$ are two different possible traces for the state of case $c$.

\begin{equation}\label{eq:PLt_1}
\footnotesize
    PL(Ad^{L^i_c},\mechanism^i) \coloneqq \underset{\substack{\simplelog^{'i},\sigma_c^i,{\sigma'}_c^i}}{sup}~log \frac{Pr(\simplelog^{'i} \mid \tilde{L}_c^i \uplus \sigma_c^i)}{Pr(\simplelog^{'i} \mid \tilde{L}_c^i \uplus {\sigma'}_c^i)}
\end{equation}
\vspace{-0.2cm}
\begin{equation}\label{eq:PLt_2}
\footnotesize
    PL(\mechanism^i) \coloneqq \underset{c \in C_L }{max}~PL(Ad^{L^i_c},\mechanism^i) 
\end{equation}

Equation~(\ref{eq:PLt_2}) is another formal representation of differential privacy that formulates the privacy budget as the supremum of privacy leakage, i.e., considering $\epsilon$ as the privacy budget, $PL(\mechanism^i) {=} \epsilon$.  
% Note that for publishing deferentially private trace variants, often a parameter is tuned to limit the length of trace variants that can be queried, e.g., the average trace length in the event log \cite{MannhardtKBWM19_short}.
% Without loss of generality, one can assume that $|\sigma_c^i|, |{\sigma'}_c^i| \leq m$, where $m$ is a large integer value.

\subsection{Privacy Leakage of Continuous Releases}\label{subsec:leakage_con}

We exploit a full-history transition system to calculate probabilities of visiting states and to generate \textit{forward} and \textit{backward} temporal correlations describing the probabilities for transitions between states.
We obtain a transition system form the last collected event log that needs to be published.  

\begin{definition}[State Probability]
Let $TS_{L,state_{hd}()}{=}(S,A,T)$ be a history transition system based on an event log $L$, the probability of visiting a state $s {\in} S$ is as follows: $Pr(s) {=} \nicefrac{|T'|}{|L|}$ where $T'{=}[(s_1,a,s_2) {\in} T | s_2{=}s]$. 
\end{definition}

For instance, in Fig.~\ref{fig:TS_history}, \small$Pr(S3)=\nicefrac{2}{4}$\normalsize. 
In the following, we define forward and backward temporal correlations based on scenarios S1 and S2 (see Subsection~\ref{subsec:scenarios}). 
Note that to simplify the notation, we abbreviate $\langle a_1,a_2,...,a_n \rangle$ as $\langle \rangle_n$, and a sequence of event logs that need to be released, i.e., $L^1,L^2,...,L^\lambda$, as $L^{1 ..\lambda}$.

% Since we assume that in each release at most one activity per case is released, temporal correlations can be defined as follows. 
% Note that we may use $TS_L$ to refer to a transition system based on an event log $L$ with $state_{hd}()$ as the state representation function.

\begin{definition}[Forward Temporal Correlations - FTC]
	\label{def:FTC}
	 Let $TS_{L,state_{hd}()}\\=(S,A,T)$ be a transition system based on an event log $L$.
	 The forward temporal correlations are calculated based on the correlations between adjacent states. Given $s_1,s_2 {\in} S$ as two adjacent states,
	 \small$Pr(s_2{=}\langle \rangle_n|s_1{=}\langle \rangle_{n-1}) = \frac{|T''|}{|T'|}$\normalsize~where \small$T' = [ (s,a,s') {\in} T \mid s{=}s_1 ]$\normalsize~and \small$T'' = [ (s,a,s') {\in} T \mid s{=} s_1 \wedge s'{=} s_2 ]$\normalsize.
% 	 \vspace{-0.1cm}
	 \begin{itemize}
	 \footnotesize
	     \item  The certain scenario with $x {\in} \mathbb{N}_{>0}$ new events:\\ 
	        Given $s_1 {\in} S {\setminus} S^{end}$, for all $s_2 {\in} S$, s.t., $s_1 {\sqsubset} s_2$, and $|s_2| - |s_1| = x$: $Pr(s_2=\langle \rangle_n | s_1=\langle \rangle_{n-x}) = \prod_{j=0}^{x-1} Pr(s_2' = \langle \rangle_{n-j} | s_1'{=}\langle \rangle_{n-(j+1)})$. Otherwise, $Pr(s_2 | s_1) = 0$.
	        If $s_1 {\in} S^{end}$, $Pr(s_2{=}s_1|s_1)=1$.
	         
	     \item The uncertain scenario with up to $x {\in} \mathbb{N}_{>0}$ new events:\\
	        Given $s_1 {\in} S {\setminus} S^{end}$, let $fd_{s_1}$ be the distance of the furthest state $s_2$ from $s_1$, s.t., $s_1 {\sqsubset} s_2$. $fm_{s_1}^{x} {=} min(x,fd_{s_1})$ is considered as the maximal forward move on the transition system starting from $s_1$. 
	        For all $s_2 \in S$, s.t., $s_1 {\sqsubset} s_2$, and for all $y \in [1,min(x,|s_2|-|s_1|)]$: $Pr(s_2=\langle \rangle_n | s_1=\langle \rangle_{n-y}) = \nicefrac{1}{(fm_{s_1}^{x}+1)} \times \prod_{j=0}^{y-1} Pr(s_2' = \langle \rangle_{n-j} | s_1'=\langle \rangle_{n-(j+1)})$, and for $y = 0$: $Pr(s_2=\langle \rangle_n | s_1=\langle \rangle_{n-y}) = \nicefrac{1}{(fm_{s_1}^{x}+1)}$. Otherwise, $Pr(s_2 | s_1) = 0$.
	        If $s_1 {\in} S^{end}$, $Pr(s_2{=}s_1|s_1)=1$.
	       %Given $s_1 {\in} S$, let $max_{ex}^{s_1,x}= min(x, max_{{s_2 \in S, s_1 \sqsubset s_2}}(|s_2| - |s_1|))$ be the maximal forward exploration depth on the transition system starting from $s_1$. 
	       %For all $s_2 {\in} S$, s.t., $s_1 {\sqsubset} s_2$, and for all ${y \in [1,min(x,|s_2|-|s_1|)]}$: $Pr(s_2=\langle \rangle_n | s_1=\langle \rangle_{n-y}) = \nicefrac{1}{(max_{ex}^{s_1,x}+1)} \times \prod_{j=0}^{y-1} Pr(s_2' = \langle \rangle_{n-j} | s_1'=\langle \rangle_{n-(j+1)})$, and for $y {=} 0$: $Pr(s_2=\langle \rangle_n | s_1=\langle \rangle_{n-y}) = \nicefrac{1}{(max_{ex}^{s_1,x})+1}$. Otherwise, $Pr(s_2 | s_1) = 0$.
	 \end{itemize}
\end{definition}

For instance, in Fig.~\ref{fig:TS_history}, given the certain scenario with $x {=} 2$, we only consider the states that their distance from a given state is 2. If the given state is \small$S1$\normalsize, \small$Pr(S5|S1) {=} \nicefrac{1}{2}$\normalsize~and \small$Pr(S6|S1) {=} \nicefrac{1}{2}$\normalsize. Other probabilities given $S1$ are considered to be zero. 
However, given the uncertain scenario with $x {=} 2$, we explore all the states within the maximal distance 2. $fd_{S1} {=} 4$ and $fm_{S1}^x {=} min (2,4)$. Thus, \small$Pr(S5|S1) {=} \nicefrac{1}{3} {\times} \nicefrac{1}{2}$\normalsize, \small$Pr(S6|S1) {=} \nicefrac{1}{3} {\times} \nicefrac{1}{2}$\normalsize, \small$Pr(S3|S1) {=} \nicefrac{1}{3} {\times} 1$\normalsize, and \small$Pr(S1|S1) {=} \nicefrac{1}{3}$\normalsize. 
Other probabilities given $S1$ are considered to be zero.
Note that in the uncertain scenario, we consider an equal chance for a case to stay in the same state, or move forward up to maximal $x$ states. This is the reason for the division by $fm_{s_1}^{x}{+}1$.

\begin{definition}[Backward Temporal Correlations - BTC]
	\label{def:BTC}
	 Let $TS_{L,state_{hd}()}\\{=}(S,A,T)$ be a transition system based on an event log $L$.
	 The backward temporal correlations can be obtained using Bayesian inference based on FTC.
	  \vspace{-0.1cm}
	 \begin{itemize}
	 \footnotesize
	    \item  The certain scenario with $x {\in} \mathbb{N}_{>0}$ new events:\\ 
	        Given $s_2 {\in} S$, for all $s_1 {\in} S$, s.t., $s_1 {\sqsubset} s_2$, and $|s_2| - |s_1| = x$: 
	        $Pr(s_1=\langle \rangle_{n-x} | s_2=\langle \rangle_{n}) = 
	        \nicefrac{Pr(s_1=\langle \rangle_{n-x}) \times Pr(s_2=\langle \rangle_{n} | s_1=\langle \rangle_{n-x})}{ Pr(s_2=\langle \rangle_{n})}$. Otherwise, $Pr(s_1 | s_2) = 0$.
	     \item The uncertain scenario with up to $x {\in} \mathbb{N}_{>0}$ new events:\\ 
	        Given $s_2 {\in} S$, let $bd_{s_2}$ be the distance of the furthest state $s_1$ from $s_2$, s.t., $s_1 {\sqsubset} s_2$, and let $bm_{s_2}^{x}= min(x,bd_{s_2})$ be the maximal backward move on the transition system starting from $s_2$. 
	       For all $s_1 {\in} S$, s.t., $s_1 {\sqsubset} s_2$, and for all ${y {\in} [1,min(x,|s_2|-|s_1|)]}$: $Pr(s_1=\langle \rangle_{n-y} | s_2=\langle \rangle_{n}) = \nicefrac{1}{(bm_{s_2}^{x}+1)} \times \nicefrac{Pr(s_1=\langle \rangle_{n-x}) \times Pr(s_2=\langle \rangle_{n} | s_1=\langle \rangle_{n-x})}{ Pr(s_2=\langle \rangle_{n})}$, and for $y {=} 0$: $Pr(s_1=\langle \rangle_{n-y} | s_1=\langle \rangle_{n}) = \nicefrac{1}{(bm_{s_2}^{x}+1)}$. Otherwise, $Pr(s_1 | s_2) = 0$.
	 \end{itemize}
\end{definition}

% Due to the tree structure of the prefix automaton, in case of certain scenarios, if $s_1$ is a prefix of $s_2$ and $|s_2| - |s_1| = x$, then $(s_1=\langle \rangle_{n-x} | s_2=\langle \rangle_{n}) = 1$. 
For instance, in Fig.~\ref{fig:TS_history}, given the certain scenario with $x{=}2$, \small  $Pr(S1|S5) = \frac{\nicefrac{2}{4} {\times} \nicefrac{1}{2} }{\nicefrac{1}{4}}$\normalsize, and given the uncertain scenario with $x{=}2$, \small$Pr(S1|S5) {=} \nicefrac{1}{3} {\times} \frac{\nicefrac{2}{4} {\times} \nicefrac{1}{2} }{\nicefrac{1}{4}}$\normalsize.
In the uncertain scenario, the previous state of a case can be the current state or any state within the maximal $x$ distance, and this is the reason for the division by $bm_{s_2}^{x} {+} 1$.
Note that we incrementally update the transition system based on the last collected event log. Thus, the knowledge regarding correlations is gained based on all the available data up to the last release point. 

% \textbf{Backward temporal correlations} can be obtained using Bayesian inference, i.e., $Pr(\sigma_c^{i-1}|\sigma_c^i)= \nicefrac{Pr(\sigma_c^{i}|\sigma_c^{i-1}) \times Pr(\sigma_c^{i-1})}{Pr(\sigma_c^{i})}$.  

% Recall that temporal correlations can be achieved based on domain knowledge as well. For example, in a heart surgery department of a hospital, the adversary could acquire the knowledge that the activity \enquote{surgery} has to follow the sequence of activities $\langle$\enquote{registration}, \enquote{bloodtest}, \enquote{hospitalization}$\rangle$.

\begin{definition}[Adversary with Temporal Correlations - $Ad^{L^{1..\lambda}_c}$]
Let $L^{1 ..\lambda}$ be the sequence of event logs that need to be released. 
We denote $Ad^{L^{1..\lambda}_c}$ as an adversary who has knowledge of all case's states in the entire releases range from 1 to $\lambda$ except the state of the victim case $c {\in} C_L$. 
The background knowledge of such an adversary is \small$L^{1..\lambda}_c = \bigcup_{i \in [1,\lambda]}L^i_c$\normalsize~as well as the knowledge of temporal correlations.
\small$Ad^{L^{1..\lambda}_c}_F (Ad^{L^{1..\lambda}_c}_B)$\normalsize~denotes such an adversary with only forward (backward) temporal correlations.
\end{definition}

% We denote $Ad_{c}^{\lambda}$ as an adversary who has knowledge of all case's states in the entire release range from 1 to $\lambda$ except the state of the victim case $c$. 
% % Thus, the background knowledge at each release point $i {\in} [1,\lambda]$ is $bk^i=L^i\setminus \{(c,\sigma) {\in} L^i \mid s_c^i{=}\pi_{\sigma}((c,\sigma)) \}$. 
% Such an adversary has also knowledge of forward and backward temporal correlations. $Ad_{c}^{\lambda,F}$ ($Ad_{c}^{\lambda,B}$) denotes such an adversary with only forward (backward) temporal correlations. 

Given $c {\in} C_L$, $Ad^{L^{1..\lambda}_c}$ observes the differentially private outputs $\simplelog^{'1}, \simplelog^{'2}, \dots, \simplelog^{'\lambda}$ of the DP mechanism $\mechanism^i$ applied to $\tilde{L}^i$ at each release point $i {\in} [1,\lambda]$ and attempts to identify the state of the case $c$.

\begin{definition}[Temporal Privacy Leakage - TPL]
	\label{def:TPL}
	Let $Ad^{L^{1..\lambda}_c}$ be an adversary with the knowledge of temporal correlations,
	$\mechanism^i$ be a DP mechanism that is applied to each event log $\tilde{L}^i$, $i {\in} [1,\lambda]$,
	and $\simplelog^{'i} {\in} rng(\mechanism^i)$ be the corresponding differentially private release at each release point.
	Considering $\sigma_c^i,{\sigma'}_c^i {\in} A_L^*$ as two different possible states for case $c {\in} C_L$, temporal privacy leakage of $\mechanism^i$ w.r.t. $Ad^{L^{1..\lambda}_c}$ is defined as follows:

	\vspace{-0.25cm}
	\begin{equation}\label{eq:TPL_1}
	\footnotesize
	    TPL(Ad^{L^{1..\lambda}_c},\mechanism^i) \coloneqq \underset{\substack{\simplelog^{'1},\dots,\simplelog^{'\lambda},\sigma_c^i,{\sigma'}_c^i }}{sup}~log \frac{Pr(\simplelog^{'1},\dots,\simplelog^{'\lambda} \mid \tilde{L}_c^i \uplus \sigma_c^i)}{Pr(\simplelog^{'1},\dots,\simplelog^{'\lambda} \mid \tilde{L}_c^i \uplus {\sigma'}_c^i)}
	\end{equation}
	\vspace{-0.2cm}
	\begin{equation}\label{eq:TPL_2}
	\footnotesize
	    TPL(\mechanism^i) \coloneqq \underset{c \in C_L}{max}~TPL(Ad^{L^{1..\lambda}_c},\mechanism^i) 
	\end{equation}
	
\end{definition}

The above-defined temporal privacy leakage can be broken down into backward and forward privacy leakages, as defined in Definition~\ref{def:BPL} and Definition~\ref{def:FPL}.

\begin{definition}[Backward Privacy Leakage - BPL] \label{def:BPL}
Backward privacy leakage of $\mechanism^i$, $i {\in} [1,\lambda]$, w.r.t. $Ad^{L^{1..\lambda}_c}_B$ is defined as follows:

\begin{equation}\label{eq:BPL_1}
	\footnotesize
	     BPL(Ad^{L^{1..\lambda}_c}_B,\mechanism^i) \coloneqq \underset{\substack{\simplelog^{'1},\dots,\simplelog^{'i},\sigma_c^i,{\sigma'}_c^i}}{sup}~log \frac{Pr(\simplelog^{'1},\dots,\simplelog^{'i} \mid \tilde{L}_c^i \uplus \sigma_c^i)}{Pr(\simplelog^{'1},\dots,\simplelog^{'i} \mid \tilde{L}_c^i \uplus {\sigma'}_c^i)}
\end{equation}
\vspace{-0.2cm}
\begin{equation}\label{eq:BPL_2}
	\footnotesize
	    BPL(\mechanism^i) \coloneqq \underset{c \in C_L}{max}~BPL(Ad^{L^{1..\lambda}_c}_B,\mechanism^i)
\end{equation}
\end{definition}

\begin{definition}[Forward Privacy Leakage - FPL] \label{def:FPL}
Forward privacy leakage of $\mechanism^i$, $i {\in} [1,\lambda]$, w.r.t. $Ad^{L^{1..\lambda}_c}_F$ is defined as follows:

\begin{equation}\label{eq:FPL_1}
	\footnotesize
	    FPL(Ad^{L^{1..\lambda}_c}_F,\mechanism^i) \coloneqq \underset{\substack{\simplelog^{'i},\dots,\simplelog^{'\lambda},\sigma_c^i,{\sigma'}_c^i}}{sup}~log \frac{Pr(\simplelog^{'i},\dots,\simplelog^{'\lambda} \mid \tilde{L}_c^i \uplus \sigma_c^i)}{Pr(\simplelog^{'i},\dots,\simplelog^{'\lambda} \mid \tilde{L}_c^i \uplus {\sigma'}_c^i)}
\end{equation}
\vspace{-0.2cm}
\begin{equation}\label{eq:FPL_2}
	\footnotesize
	     FPL(\mechanism^i) \coloneqq \underset{c \in C_L}{max}~FPL(Ad^{L^{1..\lambda}_c}_F,\mechanism^i)
\end{equation}
\end{definition}

From Equations (\ref{eq:TPL_2}), (\ref{eq:BPL_2}), and (\ref{eq:FPL_2}), we can conclude Equation (\ref{eq:TPL_sum}), which shows that to quantify the temporal privacy leakage, we need to analyze $BPL$ and $FPL$. 
% Recall that $PL(\mechanism^i)$ is the supremum of privacy leakage of a single release at release point $i$.
We subtract $PL(\mechanism^i)$ because it is included in both $BPL$ and $FPL$.

\begin{equation}\label{eq:TPL_sum}
	\footnotesize
	    TPL(\mechanism^i) = BPL(\mechanism^i) + FPL(\mechanism^i) - PL(\mechanism^i)
	\end{equation}

% \subsection{Quantifying Temporal Privacy Leakage}\label{sec:quantify_TPL}
Equations~(\ref{eq:BPL_1}) and (\ref{eq:FPL_1}) can be expanded based on Bayesian theorem to calculate backward and forward privacy leakages.

\vspace{-0.25cm}
\subsubsection{Quantifying BPL}\label{subsubsec:BPL}
As shown in~\cite{QDPContinuous_short}, using Bayesian theorem, $BPL(Ad^{L^{1..\lambda}_c}_B,\mechanism^i)$ can be simplified as Eq.~(\ref{eq:BPL}) (cf. Theorem 2 and Eq. (12) in \cite{QDPContinuous_short}). 
Since CEDP is incremental, the trace of a case at release point $i{-}1$ cannot be longer than its trace at release point $i$. Thus, $A^{\leq}_{\sigma} = \{ \sigma' {\in} A_L^* \mid |\sigma'| \leq |\sigma| \}$ is the domain of all possible previous steps.

\vspace{-0.15cm}

\scriptsize
\begin{flalign}\label{eq:BPL}
    BPL(Ad^{L^{1..\lambda}_c}_B,\mechanism^i) & {=} \underset{\substack{\simplelog^{'1},\dots,\simplelog^{'{i-1}} \\ 
    \sigma_c^i,{\sigma'}_c^i}}
    {sup}~log \frac{\sum\limits_{\sigma_c^{i-1} {\in} A^{\leq}_{\sigma_c^i}}Pr(\simplelog^{'1},\dots,\simplelog^{'{i-1}} \mid \tilde{L}^{i-1}_c \uplus \sigma_c^{i-1})Pr(\sigma_c^{i-1}|\sigma_c^i)}{\sum\limits_{{\sigma'}_c^{i-1} {\in} A^{\leq}_{{\sigma'}_c^i} }\underbrace{Pr(\simplelog^{'1},\dots,\simplelog^{'{i-1}} \mid \tilde{L}^{i-1}_c \uplus {\sigma'}_c^{i-1})}_{\mathclap{(a)}}\underbrace{Pr({\sigma'}_c^{{i-1}}|{\sigma'}_c^{i})}_{\mathclap{(b)}}} &&\\ \nonumber
    & + \underset{\substack{\simplelog^{'i},\sigma_c^i,{\sigma'}_c^i}}{sup}~log \frac{Pr(\simplelog^{'i} \mid \tilde{L}_c^i \uplus \sigma_c^i)}{\underbrace{Pr(\simplelog^{'i} \mid \tilde{L}_c^i \uplus {\sigma'}_c^i)}_{\mathclap{(c)}}} &&
\end{flalign}
\normalsize

In Eq.~(\ref{eq:BPL}), the part annotated with (a) refers to BPL at point $i{-}1$, (b) refers to the backward conditional probabilities for the case $c$ given its state at release point $i$, and (c) is the privacy leakage of single release at point $i$ without considering temporal correlations. Based on Eq.~(\ref{eq:BPL}), if $i{=}1$, then $BPL(Ad^{L^{1..\lambda}_c}_B,\mechanism^1) {=} PL(Ad^{L_c^1},\mechanism^1)$, and if $i {>} 1$, $BPL(Ad^{L^{1..\lambda}_c}_B,\mechanism^i)$ is as follows, where $AL_B(.)$ is a function to calculate the accumulated BPL.

\begin{equation}\label{eq:BPL_AL}
\footnotesize
	BPL(Ad^{L^{1..\lambda}_c}_B,\mechanism^i) {=} AL_B(BPL(Ad^{L^{1..\lambda}_c}_B,\mechanism^{i-1})) + PL(Ad^{L^i_c},\mechanism^i)
\end{equation}

Equation~(\ref{eq:BPL_AL}) shows that BPL can be calculated recursively and may accumulate over time. According to Definition~\ref{def:BTC}, and considering the certain scenario of event data publishing, the backward temporal correlation between states of a case is always on an extreme side, i.e., given $\sigma_c^i$ as the state of a case $c {\in} C_L$ at the release point $i {\in} [1,\lambda]$, there exists a state for the case $c$ at release point $i{-}x$, $\sigma_c^{i-x}$, s.t., $\sigma_c^{i-x} {\sqsubset} \sigma_c^{i}$, thus $Pr(\sigma_c^{i-x}|\sigma_c^i){=}1$. Consequently, considering $i{=}2$, $BPL(Ad^{L^{1..\lambda}_c}_B,\mechanism^2)$ is calculated as follows:

\vspace{-0.1cm}
\scriptsize
\[
BPL(Ad^{L^{1..\lambda}_c}_B,\mechanism^2) = 
\underset{\simplelog^{'1},\sigma_c^1,{\sigma'}_c^1}{sup}~log \frac{Pr(\simplelog^{'1} \mid \tilde{L}^1_c \uplus \sigma_c^1)}{Pr(\simplelog^{'1} \mid \tilde{L}^1_c \uplus {\sigma'}_c^1)} +
\underset{\simplelog^{'2},\sigma_c^2,{\sigma'}_c^2}{sup}~log \frac{Pr(\simplelog^{'2} \mid \tilde{L}^2_c \uplus \sigma_c^2)}{Pr(\simplelog^{'2} \mid \tilde{L}^2_c \uplus {\sigma'}_c^2)}
\]
\vspace{-0.2cm}
\[
= PL(Ad^{L^1_c},\mechanism^1) + PL(Ad^{L^2_c},\mechanism^2)
\]
\normalsize

If we consider $\epsilon$ as the privacy budget of the mechanism $\mechanism^i$, i.e., for any $i \in [1,\lambda]$, ${max}_{c \in C_L}PL(Ad^{L_c^i},\mechanism^i) {=} \epsilon$. Then, $BPL(Ad^{L^{1..\lambda}_c}_B,\mechanism^2) {=} 2\epsilon$.
Consequently, $BPL(Ad^{L^{1..\lambda}_c}_B,\mechanism^3) {=} 3\epsilon$, $BPL(Ad^{L^{1..\lambda}_c}_B,\mechanism^4) {=} 4\epsilon$, etc. Hence, BPL for CEDP considering the certain scenario is expected to linearly increase. We investigate this observation in our experiments.

\subsubsection{Quantifying FPL}\label{subsubsec:FPL}
Similar to the backward privacy leakage, the equation of the forward privacy leakage, i.e., Eq.~(\ref{eq:FPL_1}), can also be simplified as Eq.~(\ref{eq:FPL}) (cf. Theorem 2 and Eq. (14) in \cite{QDPContinuous_short}). 
Since continuous event data publishing is incremental, the trace of a case at release point $i+1$ cannot be shorter than its trace at release point $i$. Thus, $A^{\geq}_{\sigma} = \{ \sigma' \in A_L^* \mid |\sigma| \leq |\sigma'| \}$.

\vspace{-0.1cm}

\scriptsize
\begin{flalign}\label{eq:FPL}
    FPL(Ad^{L^{1..\lambda}_c}_F,\mechanism^i) & {=} \underset{\substack{\simplelog^{'{i+1}},\dots,\simplelog^{'\lambda} \\ 
    \sigma_c^i,{\sigma'}_c^i}}
    {sup}~log \frac{\sum\limits_{\sigma_c^{i+1} {\in} A^{\geq}_{\sigma_c^i}}Pr(\simplelog^{'{i+1}},\dots,\simplelog^{'\lambda} \mid \tilde{L}^{i+1}_c \uplus \sigma_c^{i+1})Pr(\sigma_c^{i+1}|\sigma_c^i)}{\sum\limits_{{\sigma'}_c^{i+1} {\in} A^{\geq}_{{\sigma'}_c^i} }\underbrace{Pr(\simplelog^{'{i+1}},\dots,\simplelog^{'\lambda} \mid \tilde{L}^{i+1}_c \uplus {\sigma'}_c^{i+1})}_{\mathclap{(a)}}\underbrace{Pr({\sigma'}_c^{{i+1}}|{\sigma'}_c^{i})}_{\mathclap{(b)}}} &&\\ \nonumber
    & + \underset{\substack{\simplelog^{'i},\sigma_c^i,{\sigma'}_c^i}}{sup}~log \frac{Pr(\simplelog^{'i} \mid \tilde{L}_c^i \uplus \sigma_c^i)}{\underbrace{Pr(\simplelog^{'i} \mid \tilde{L}_c^i \uplus {\sigma'}_c^i)}_{\mathclap{(c)}}} &&
\end{flalign}
\normalsize

In Eq.~(\ref{eq:FPL}), the part annotated with (a) refers to FPL at release point $i+1$, (b) refers to the forward conditional probabilities for the case $c$ given its state at point $i$, and (c) is the privacy leakage of single release at point $i$ without considering temporal correlations. Similar to Eq.~(\ref{eq:BPL}), in Eq.~(\ref{eq:FPL}), if $i=1$, then $FPL(Ad^{L^{1..\lambda}_c}_F,\mechanism^1) {=} PL(Ad^{L_c^1},\mechanism^1)$, and if $i {>} 1$, $FPL(Ad^{L^{1..\lambda}_c}_F,\mechanism^i)$ is as follows, where $AL_F(.)$ is a function to calculate the accumulated forward privacy leakage.

\begin{equation}\label{eq:FPL_AL}
\footnotesize
	FPL(Ad^{L^{1..\lambda}_c}_F,\mechanism^i) {=} AL_F(FPL(Ad^{L^{1..\lambda}_c}_F,\mechanism^{i+1})) + PL(Ad^{L_c^i},\mechanism^i)
\end{equation}

Equation~(\ref{eq:FPL_AL}) shows that FPL can also be recursively calculated and may accumulate over time. Since event data publishing is incremental, the complete traces remain the same in all the next releases. Thus, FPL in CEDP can be on an extreme side whenever there exist complete traces in the previous releases. 
Moreover, based on Eq.~(\ref{eq:FPL}), we assume FPL in CEDP depends on the variation of traces in an event log.  
For instance, considering the certain scenario, if an event log only contains one trace variant, FPL can be on an extreme side because the next state of a new case is certainly known based on the previously recorded states. Hence, FPL is expected to linearly increase. We investigate the effect of the \textit{trace uniqueness} ratio on FPL in our experiments.

% In contrast, if all the cases have unique traces in an event log, most of the states obtained from the last collected event log are new, and one cannot gain knowledge from the transition system of the previously recorded states. Thus, FPL is expected to remain unchanged. We investigate these observations in our experiments.

% Note that based on Definition~\ref{def:FTC}, for each case, the probability of visiting a state at the release point $i{+}1$, given its state at the point $i$, can only be calculated if a corresponding transition has already been seen in the previously collected event logs. Otherwise, we consider the worst-case w.r.t. the knowledge gained based on the correlations, i.e., no correlation.

\subsubsection{Calculating Accumulative Privacy Leakage}\label{subsubsec:cumulative_PL}
Cao et al.~\cite{QDPContinuous_short} show that the accumulative privacy leakages can be formulated as an optimization problem where the objective function is a ratio of two linear functions and the constraints are linear equations. 
Since we rely on transition systems to obtain temporal correlations, the knowledge of temporal correlations is bounded to the traces in the state space of the transition system. 
For the traces that are not included in the state space, we consider the worst-case w.r.t. the knowledge of correlations, i.e., no correlation. 
We assume that adapting the optimization problem from \cite{QDPContinuous_short}, is a straightforward process. Thus, we avoid including it here. Nevertheless, we provided the adapted optimization problem and a short explanation regarding the \textit{computational complexity} of our approach as supplementary material in our GitLab repository.\footnote{\scriptsize\url{https://github.com/m4jidRafiei/QDP_CEDP/tree/main/supplementary}}

\section{Experiments}
\label{sec:experiments}

The aims of the experiments are as follows: (1) Investigating the effect of temporal correlations among event logs on the provided privacy guarantees, (2) Exploring the effect of different CEDP scenarios on temporal correlations and privacy leakages, and (3) Exploring the impact of trace uniqueness in event logs on temporal privacy leakages.
We have implemented a Python script to conduct the experiments. The source code is available on GitLab\footnote{\scriptsize\url{https://github.com/m4jidRafiei/QDP_CEDP}} and as a Python package\footnote{\scriptsize\url{https://pypi.org/project/pm-cedp-qdp/}} that can be installed using \textit{pip} commands.
Table~\ref{tbl:eventlogs_exp} shows the general statistics of the real-life public event logs that we employed for our experiments. The \textit{trace uniqueness} shows the rate of unique traces, i.e., $\nicefrac{\#Variants}{\#Traces}$. These event logs cover a wide range w.r.t. the trace uniqueness.
\begin{table}[t]
\scriptsize
\centering
\caption{General statistics of the event logs used in the experiments.}\label{tbl:eventlogs_exp}
\begin{tabular}{@{}l|l|l|l|l|l@{}}
\hline
Event Log     & \#Events & \#Unique Activities & \#Traces & \#Variants & Trace Uniqueness \\ \hline
Sepsis  & 15214     & 16                   & 1050     & 846         & 80\%           \\ 
BPIC-2013     & 65533     & 4                    & 7554     & 1511        & 20\%           \\
BPIC-2012-App & 60849     & 10                   & 13087    & 17          & 0.12\%         \\ \hline
\end{tabular}
\end{table}

\begin{table}[b]
\scriptsize
\centering
\vspace{-0.4cm}
\caption{General statistics of the initial and continuous parts of event logs used in the experiments.}\label{tbl:eventlogs_split}
\begin{tabular}{@{}l|l|l|l|l@{}}
\hline
Event Log     & Parts      & \#Events & \#Complete Traces & \#Incomplete Traces \\ \hline
Sepsis  & Initial    & 7290     & 442              & 84                \\ \cmidrule(l){2-5} 
              & Continuous & 7924        & 524                 & 84                  \\ \hline
BPIC-2013     & Initial    & 21705     & 0              & 2271                \\ \cmidrule(l){2-5} 
              & Continuous & 43828     & 5283                 & 2271                \\ \hline
BPIC-2012-App & Initial    & 29227     & 5849             & 690               \\ \cmidrule(l){2-5} 
              & Continuous & 31622      & 6548                 & 690                 \\ \hline
\end{tabular}
\end{table}

To simulate CEDP, we need to specify the initial release and a sequence of event logs that are considered to be continuously published. Thus, we need a \textit{split-point} that splits an event log into two parts; \textit{initial} and \textit{continuous}. 
One can partition an event log into initial and continuous parts in a variety of ways, e.g., having an initial log that contains all the cases, or having an initial log that contains $x$\% of cases or events, and so on. 
We consider the percentage of events included in the initial part as a criteria for splitting an event log.
We split Sepsis and BPIC-2012-App into two parts such that the initial part contains roughly 50\% of events so that there is enough data to obtain reliable knowledge regarding the correlations. However, BPIC-2013 is partitioned in such a way that the initial part contains roughly 35\% of events so that there exist no complete trace, yet, at the same time, there is enough data to discover a transition system and obtain the probabilities.\footnote{\scriptsize Note that experiments can be extended considering different partitioning scenarios and focusing on different log characteristics.}
% For our experiments, we consider roughly 50\% of events to be included in the initial part.
Table~\ref{tbl:eventlogs_split} shows general statistics of the event logs partitions after being partitioned. 
Note that incomplete (partial) traces are the same in both partitions.

The initial part is published as the first release. Then, each future release is generated w.r.t. the scenarios S1 and S2 (see Subsection~\ref{subsec:scenarios}).
In both scenarios, the window size, i.e., the number of new events per trace in a future release, varies from 1 to 4.
Note that to simulate scenario S2, a random integer within the window size is generated to determine the number of new events.
For each scenario, we continue the publishing process for up to 5 releases or until there are no incomplete traces to publish. 

% \begin{figure}[t]
% 	\centering
% 	\subfloat[\scriptsize Sepsis. ]{\includegraphics[width=0.49\textwidth]{figures/Sepsis-200.jpg}\label{fig:sepsis-200}}
% 	\hfill
% 	\subfloat[\scriptsize BPIC-2012-App.
% 	]{\includegraphics[width=0.49\textwidth]{figures/BPIC2012App-200.jpg}\label{fig:BPIC2012-app-200}} \hfill
% 	\subfloat[\scriptsize BPIC-2013. 
% 	]{\includegraphics[width=0.70\textwidth]{figures/BPIC2013-200.jpg}\label{fig:BPIC2013-200}} \hfill
	
% 	\caption{The forward privacy leakage (FPL), the backward privacy leakage (BPL), and overall temporal privacy leakage (TPL) for different releases of the event logs. The full-history of traces is considered as the state representation function and $\epsilon=0.01$.}
% 	\label{fig:privacy_leakages_200}
% 	\vspace{-0.4cm}
% \end{figure}

\begin{figure}[t]
	\centering
	\includegraphics[width=0.99\textwidth]{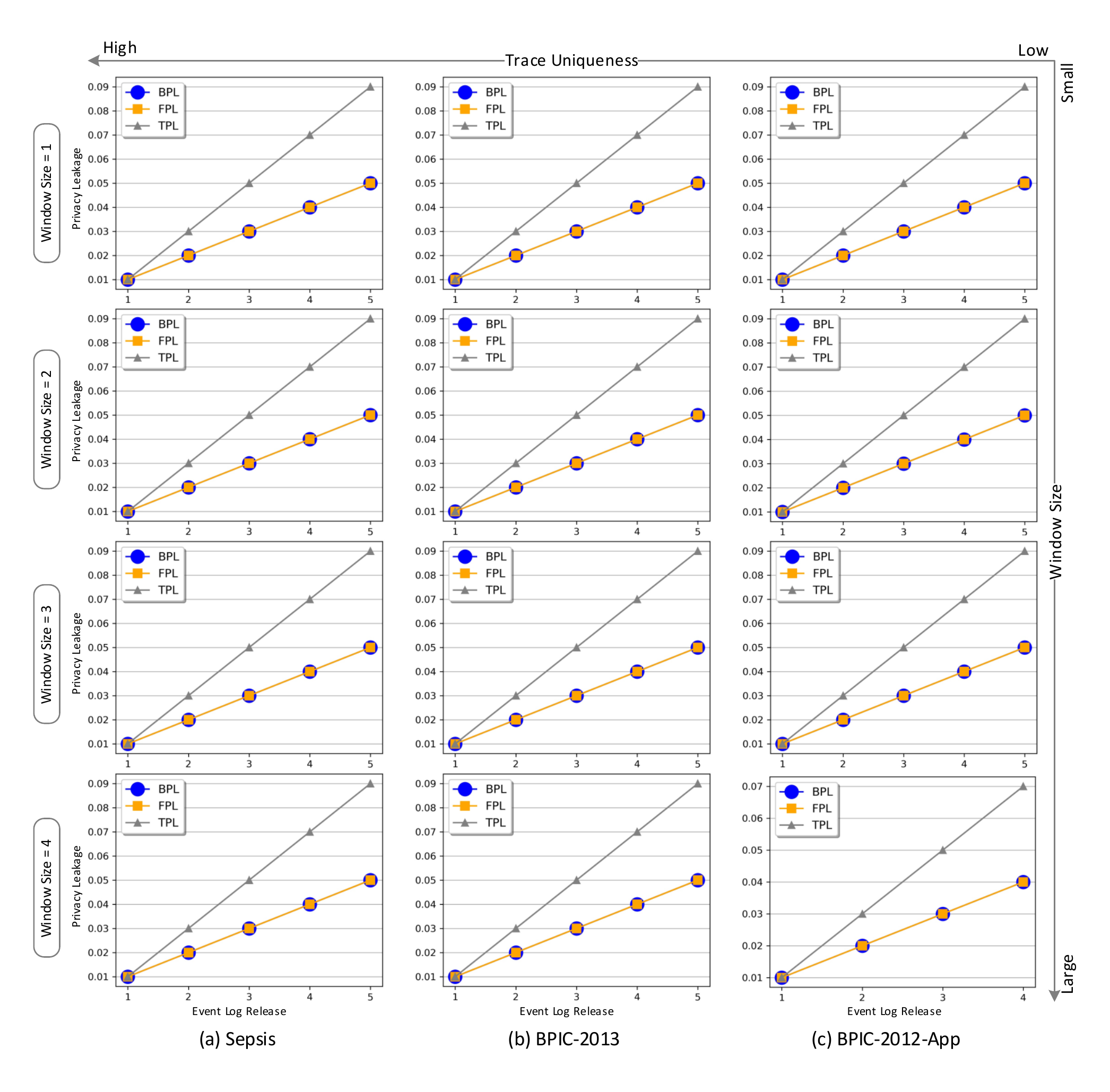}
	\caption{FPL, BPL, and TPL for different releases of the event logs, when the CEDP scenario is S1, window size varies from 1 to 4, and  $\epsilon=0.01$. The window size indicates the number of new events per trace in a future release.}
	\label{fig:S1}
	\vspace{-0.5cm}
\end{figure}

\begin{figure}[t]
	\centering
	\includegraphics[width=0.99\textwidth]{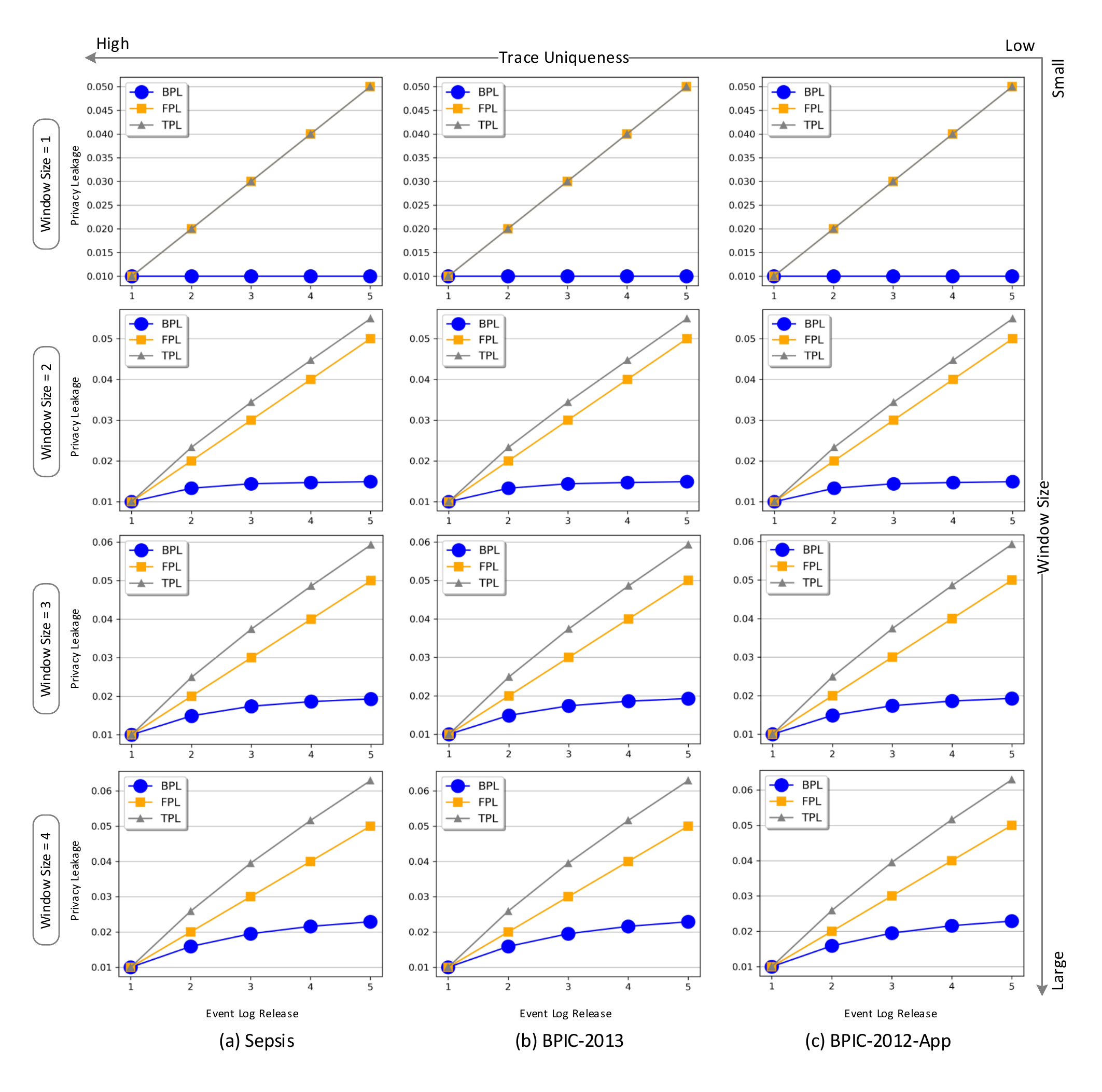}
	\caption{FPL, BPL, and TPL for different releases of the event logs, when the CEDP scenario is S2, window size varies from 1 to 4, and  $\epsilon=0.01$. The window size indicates the maximum number of new events per trace in a future release.}
	\label{fig:S2}
	\vspace{-0.6cm}
\end{figure}

Figure~\ref{fig:S1} and \ref{fig:S2} show the privacy leakages for different releases of the event logs based on the CEDP scenarios S1 and S2, respectively.
We consider $\epsilon=0.01$ as the privacy budget of a differential privacy mechanism $\mechanism$ that is applied to each release. 
Thus, for the first release \small$FPL{=}BPL{=}TPL{=}0.01$\normalsize. Recall that \small$TPL{=}FPL{+}BPL{-}\epsilon$\normalsize. 
Note that the implementation details of such a mechanism that does not consider correlations among different releases will not impact our experiments. In the following, we explain the results for each scenario.

\textbf{Scenario S1:} The first observation is that the results are the same for all the event logs. The only different plot is at the bottom right with less number of releases because there exists no incomplete trace for BPIC-2012-App after the 4th release. 
Since the previous states are certain, for each state there is one state with $BTC=1$. Thus, the correlations are strong, and BPL linearly increases for all the event logs.
The same results can be seen for FPL due to different reasons. 
In Sepsis and BPIC-2012-App, FPL linearly increases because initial releases of these event logs contain complete traces that remain unchanged in all the next releases. Thus, there are strong correlations among those traces in all the releases. Moreover, for the most of the incomplete states (traces) there exist certain states in future releases. For example, in the second release of Sepsis, almost 78\% of the incomplete states have a certain state when window size is 2.
We see the same trend in BPIC-2013 although there exist no complete trace in its initial release. This is because of two reasons: (1) there are complete traces that appear in the second release, which is used to discover the updated transition system, and (2) BPIC-2013 contains a few distinct activities that leads to strong correlations between states. 
In BPIC-2013, there exist only 4 unique activities, and 86\% of variants contain only two activities \enquote{Accepted} and \enquote{Queued}. That leads to a situation where for many states in the corresponding transition system there exists only one possible next state that results in strong correlations.

When the window size is increased, one may expect to see lower forward correlations between states. Particularly, for the event logs with a high trace uniqueness. However, due to more complete traces that appear by increasing the window size, FPL does not decrease. We continued releasing Sepsis event logs considering 4 and 8 as window sizes until there was no more incomplete traces. According to the results, FPL never decreased.\footnote{\scriptsize\url{https://github.com/m4jidRafiei/QDP_CEDP/tree/main/more_exp}}
Moreover, the trace uniqueness that may affect FPL does not show any impact because of the existence of strong correlations between states in all the event logs.

\textbf{Scenario S2:} The first observation is that all the event logs follow the same trend based on the window size. 
One can see a logarithmic increase for BPL based on the window size that corresponds to the so-called moderate type of correlations. That is because for the larger window sizes more states are explored on the corresponding transition system. Thus, more knowledge is gained regarding correlations. However, at the same time, more uncertainty is imposed because the previous state can be any state within the window size distance (see Definition~\ref{def:BTC}).
FPL still linearly increases, similar to scenario S1, which is mainly due to the complete traces leading to strong correlations.
% Note that in contrast to scenario S1, the existence of a few distinct activities in BPIC-2013 cannot result in strong correlations in scenario S2 due to the imposed uncertainty by the data publishing scenario. 
Also, the results do not change based on the trace uniqueness because of the existence of the strong correlations.   

\textbf{Scenario S1 vs Scenario S2:}
By comparing the two CEDP scenarios, one can see that scenario S1, as a certain scenario, leads to higher privacy leakages, as expected. That is because certain scenarios result in stronger correlations. This observation shows that not revealing exact data publishing scenarios in CEDP can mitigate temporal privacy leakages to some extent.

% However, FPL behaves differently for each event log which is consistent with our analysis, i.e., it is lower (higher) for event logs with higher (lower) trace variation.
% The case uniqueness of the Sepsis event log is 80\%, i.e., high trace variation, which results in low values for $FPL$. 
% However, the case uniqueness of BPIC-2012-App is 0.12\% which results in linearly incremental $FPL$, similar to $BPL$. 
% For BPIC-2013 whose case uniqueness is 20\%, FPL linearly increases until the $15^{th}$ release and then drops to 0.01 which is the $\epsilon$ value. That means, from the $16^{th}$ release, the traces contain patterns that have not been seen in the previous releases. 
% Note that in the last release of all the event logs, $FPL{=}\epsilon$ which is due to the fact that there is no next state for completed traces. Thus, there is no correlation.

% Although event logs with higher trace variation (e.g., Sepsis) are challenging for privacy preservation techniques proposed for a single release \cite{MannhardtKBWM19_short,RafieiA21Group,SaCoFa}, our analysis and experiments show that such event logs provide less knowledge regarding temporal correlations. 
% Thus, one may need to provide stronger privacy guarantees for the first release of such event logs. However, less noise needs to be added to prevent possible temporal privacy leakages when these event logs are continuously published.  
% At the same time, one should not underestimate the noise that needs to be added to the event logs with low trace variation when they are continuously published.   

\section{Conclusion and Discussion}
\label{sec:conc_disc}
In this paper, we quantified the privacy leakage of differential privacy mechanisms in the context of continuous event data publishing under temporal correlations. We utilized transition systems to model and quantify the correlations. We did experiments on real-life public events logs considering different CEDP scenarios. Our experiments showed that privacy leakage of a differential privacy mechanism may increase over time. 
In the following, we discuss some design choices, possible next steps, and limitations that need to be taken into account. 

The concept of state, which is defined based on a state representation function, provides a general way to quantify correlations w.r.t. sensitive data. For instance, if one considers the set of activities in a trace as sensitive data rather than the sequence, and the corresponding differential privacy mechanism aims to protect the set of activities in traces. Then, our approach can be adapted to quantify the corresponding temporal privacy leakage by changing the state representation function, s.t., each state represents the set of activities in a trace.

% Second, although our correlation model, i.e., the transition system, is the same for all the cases, we still provide our formal definitions for conditional probabilities at the case level. That is because the transition system can be enriched with some case-specific information. For instance, $Pr(s_5|s_3){=}0.9$ if the age of case is $>30$ and $Pr(s_5|s_3){=}0.5$ if the age of case is $\leq30$. 

The incrementally updated transition system based on the last collected event log may not be reliable for calculating forward temporal correlations if it contains only a few states. To gain more reliable knowledge regarding the correlations, one can consider a minimum number of cases reflecting a specific correlation. One can also apply more conditions, such as only considering the correlations obtained based on complete traces.  

% Third, for the sake of simplicity, we considered publishing only one event for the incomplete traces at a release step. However, the approach can be applied based on any size of window steps and using the \textit{Kolmogorov equation} to calculate correlations, i.e., conditional probabilities.     
We only considered the control-flow aspect of event logs, while in reality the events recorded by information systems often contain more attributes.
Each event attribute in a trace can be used to create a new correlation model or to alter an existing one. Depending on the attributes present in published event logs, one may need to analyze the corresponding correlations to examine possible privacy leakages.      
Overall, this work highlights the necessity of designing differential privacy mechanisms that consider temporal correlations when event data are continuously published. 

% \section{Conclusion}
% \label{sec:conclusion}
% In this paper, we quantified the privacy leakage of differential privacy mechanisms in the context of continuous event data publishing under temporal correlations. We relied on transition systems to model and quantify the correlations based on previous observations. We did experiments on real-life events logs considering a continuous release scenario. Our experiments showed that privacy leakage of a differential privacy mechanism may increase over time. 
% In Section~\ref{sec:discussion}, we described several design choices, limitations, and possible next steps. Overall, this work highlights the necessity of designing differential privacy mechanisms that consider temporal correlations when event data are continuously published. 


\bibliographystyle{splncs04}
\bibliography{Refrences}

\end{document}